\documentclass[conference]{IEEEtran}

\usepackage{courier}
\usepackage{amstext}
\usepackage{times}
\usepackage{helvet}
\usepackage{courier}
\usepackage[cmex10]{amsmath}
\usepackage{amssymb}
\usepackage{latexsym}
\usepackage{eurosym}
\usepackage{wasysym}

\makeatletter
\newif\if@restonecol
\makeatother

\usepackage{etex}
\usepackage{tikz}
\usepackage{pslatex}
\usepackage[arrow, matrix, curve]{xy}
\usepackage{epsfig}
\usepackage{pstricks,egameps}
\usepackage{hyperref, url}

\begin{document}

\title{Experimental economics for web mining}

\author{\IEEEauthorblockN{Rustam Tagiew}
\IEEEauthorblockA{Alumni of\\TU Bergakademie Freiberg\\
yepkio@mail.ru}
\and
\IEEEauthorblockN{Dmitry I. Ignatov}
\IEEEauthorblockA{National Research University\\
Higher School of Economics\\
dignatov@hse.ru}
\and
\IEEEauthorblockN{Fadi Amroush}
\IEEEauthorblockA{Granada Lab of\\
Behavioral Economics (GLOBE)\\
Granada, Spain\\
fadyamr@gmail.com}}

\maketitle

\begin{abstract}
This paper offers a step towards research infrastructure, which makes data from experimental economics efficiently usable for analysis of web data. We believe that regularities of human behavior found in experimental data also emerge in real world web data. A format for data from experiments is suggested, which enables its publication as open data. Once standardized datasets of experiments are available on-line, web mining can take advantages from this data. Further, the questions about the order of causalities arisen from web data analysis can inspire new experiment setups. 
\end{abstract}

\IEEEpeerreviewmaketitle

\section{Introduction}
\indent From the view point of web miners or rather data scientists dealing with human behavior, the sizes of datasets originated from web predominate the ones from experimental economics by orders of magnitude \cite{vspt}. Nevertheless, analyzing data from experimental economics has the same importance for understanding human psychology as studying E. coli for understanding human physiology. It is reasonable to assume that human behavior regularities found in experimental data also emerge in web data and vise versa. Data from experimental economics has the advantage of originating from simple and controlled human interactions.\\
\indent The reason for economists to conduct experiments was the deficiency of the pure theory. It is not only the global financial crisis of the recent years \cite{maxton}, which made economists reconsider the path economics as a discipline should take. Since decades, it became obvious that classical theories fail in real world \cite{ariely}. Paul Krugman described the current situation in economics as: ``... the central cause of the profession’s failure was the desire for an all-encompassing, intellectually elegant approach that also gave economists a chance to show off their mathematical prowess. Unfortunately, this romanticized and sanitized vision of the economics led most economists to ignore all the things that can go wrong. They turned a blind eye to the limitations of human rationality''.\\
\indent Experimental economics gained its importance as a promising solution of this problem. Experiments on human subjects have shifted economists' point of view closer to the psychologists' one -- people are no more considered to be rational monetary payoff maximizers alias homines oeconomici \cite{behhomo}. This makes economic research more useful for web mining, since web users do not always interact in monetized or monetizable domains. Laboratory and field experiments are carried out using human subjects in order to improve theoretical knowledge about human behavior in interactions. The body of knowledge or rather the discipline arising from this process is called behavioral economics.\\
\indent Today, behavioral economists simplify real world cases to an experiment setup, make an assumption about the behavior model, conduct experiments, fit their model's parameters and come back with an explanation of the real world human behavior. Unfortunately, many presumed models from economics might be conceptually not appropriate to fit human behavior. For instance, linear models are commonly used on nominal data \cite{experimental, sefton}, whereby nominal value sets are arbitrarily projected into real numbers. Otherwise, non-deterministic models as QRE \cite{qre} are used to explain human behavior, although human subjects fail at generating truly random sequences and at dealing with probabilities \cite{wagenaar, nonlogic}. It is recommended to reanalyze data in some cases instead of trusting the published results. Therefore, a facility for publishing original experimental data has to be created. In fact, closing the gap between behavioral economics and data science remains a challenge for future.\\
\indent As for behavioral economics, as for web mining and as for any scientific discipline, real world applications have to be at least the long-term results. Behavioral economists and web miners are confronted with the same problem -- they have to predict human behavior and to develop policies to steer it the desired way. Starting from a real world question, behavioral economists and web miners branch into two different and mutually completing approaches. Behavioral economists can find true causalities, as long these causalities match into simple model spaces. Whereby, web miners can generate complicated models, but are not able to decide the order of causalities. Once the expert knowledge from behavioral economists, formatted experimental data, advanced machine learning algorithms and web data come together, more reliable solutions can be provided to real world problems.\\  
\section{Related Work}
\indent Regarding open experimental data, there was the by now defunct website ExLab (exlab.bus.ucf.edu) launched by the department of Economics of University of Central Florida. This website included only data from non-interactive experiments. Further, the data was not uniformly formatted.\\
\indent Market leaders already push forward into the area between behavioral economics and data science. Facebook hired economists for studying user and advertiser behavior, economic networks, incentives, externalities, and decision making under risk and uncertainty \cite{facebookbe}. Microsoft founded MSR-NYC \cite{microsoftbe}, where researchers develop technologies in the intersection of social science, both computational and behavioral, computational economics and prediction markets, machine learning, as well as information retrieval. Google's chief economist openly writes in his paper \cite{varian}: ``I believe that [manipulating and analyzing big data] have a lot to offer and should be more widely known and used by economists. In fact, my standard advice to graduate students these days is `go to the computer science department and take a class in machine learning'.''\\ 
\indent In the academic world, the international workshop series ``Experimental Economics and Machine Learning'' (EEML) started in 2012 \cite{eeml1be,eeml2be,eeml3be}. EEML seeks to fill the gap between two scientific communities of Experimental Economics and AI \& Data Mining. The conference ``Social media and behavioral economics'' took place in 2013 \cite{harvardbe}, where data scientists and economists from universities and industry participated. Yale University has a chair researching in machine learning, behavioral economics, and finance \cite{brown}.\\
\section{Data Format for Experimental Economics}
\begin{table}
\caption{$5$ independent data sets.}
\label{datasets}
\centering
\begin{tabular}{|l|r|}
\hline
Dataset                          &   Number of human decisions\\
\hline
Roshambo \cite{tagiewaiia}  & 4000 \\
Colored Trails (Responder) \cite{antospfeffer} & 371\\
Product Selection \cite{famroush} & 145 \\
Social Learning \cite{soclearnnet} &  14040\\
Gift-Exchange-Game \cite{sefton} & 746\\
\hline
\end{tabular}
\end{table}
\indent The authors of this paper have the cumulated experience on analyzing $5$ independent datasets from experimental economics \cite{tagieweeml,famroush,tagieweeml2,tagiewreci}. Tab.\ref{datasets} shows the number of single human decision in every of the datasets. As a single human decision, we define one no further indivisible turn executed by one human. For every turn, there is some information visible to the decision taking subject. In a subsequent decision, this information also includes own previous decisions and the information, which was visible earlier. Once a set of decision samples with equal structure of visible information is created, data mining can be conducted. Tab.\ref{refinedss} shows an example of a refined set of decision samples.\\
\begin{table}[b]
\caption{Example for a refined set of decision samples. David's decision is to predict.}
\label{refinedss}
\centering
\begin{tabular}{|c|c|c|c|c|}
\hline
World & Alice & Bob & Charly & \textbf{David} \\
\hline
sun & beach & beach & library & \textbf{library} \\
rain & library & cinema & library & \textbf{library} \\
cloudy & park & park & beach & \textbf{beach} \\
$\vdots$ & $\vdots$ & $\vdots$ & $\vdots$ & \textbf{$\vdots$} \\
\hline
\end{tabular}
\end{table}
\indent From our experience, the data refinery until this point made by an external data scientist takes more time as the actual data mining. Especially, determining information visible to a player is a time consuming process. Therefore, we suggest a format for the interchange between economists and data scientists, where all same structured decision samples are organized in separated sets as already exemplarily shown on Tab.\ref{refinedss}. We further call visible 
information as inputs. For instance, the ``Colored Trails'' dataset has only one set of decision samples with `own payoff' and `partner's payoff' as inputs.\\
\section{Community Formation}
\indent The issues expressed before should not reduce experimental economics to data creation, although the knowledge creation conducted solely by economists is not optimal. We rather suggest to relocate economists from the start to the end of the knowledge creation process. That means that the data is analyzed first by data scientists and then the regularities are corrected and interpreted by economists. Therefore, every research project has to include data scientists as well as economists in its participants' list. A web resource offering uniformly formatted datasets should serve as a focal point for the crystallization of such research teams. The data license is an issue to be solved for such a facility.\\
\section{Conclusion}
\indent We suggested a simple format for efficient interchange of experiment data between economists and data scientists, which should facilitate the cooperation between them to conduct research on real world problems concerning human behavior on web.\\ 
\bibliographystyle{IEEEtran}
\newpage
\bibliography{webminexpeco}
\end{document}